\documentclass{iau}
\usepackage{graphicx}
\def\simlt{\lower.5ex\hbox{$\; \buildrel < \over \sim \;$}}
\def\simgt{\lower.5ex\hbox{$\; \buildrel > \over \sim \;$}}
\def\kms{km s$^{-1}$}

\def\schi{{\sc Hi}\ }

\title[IAU 296.~~{[Fe II]} and Dust Emissions from SNRs] 
{Infrared [Fe II] and Dust Emissions from Supernova Remnants}

\author[Bon-Chul Koo]   
{Bon-Chul Koo$^1$
 }

\affiliation{$^1$Department of Physics and Astronomy, Seoul National University, \\ Seoul 151-747, KOREA
\\ email: {\tt koo@astro.snu.ac.kr}}

\pubyear{2013}
\volume{296}  
\pagerange{--}
\jname{Supernova Environmental Impacts}
\editors{Alak K. Ray eds.}
\begin{document}

\maketitle

\begin{abstract}

Supernova remnants (SNRs) are strong thermal emitters of 
infrared radiation. 
The most prominent lines in the near-infrared spectra of SNRs 
are [Fe II] lines. The [Fe II] lines are 
from shocked dense atomic gases, so they trace SNRs
in dense environments.
After briefly reviewing the physics of the [Fe II] 
emission in SNR shocks, I describe the observational results which show 
that there are two groups of SNRs bright in [Fe II] emission: 
middle-aged SNRs interacting with molecular 
clouds and young core-collapse SNRs in 
dense circumstellar medium. 
The SNRs belonging to the former group are also  
bright in near-infrared H$_2$ emission, indicating that 
both atomic and molecular shocks are pervasive in these SNRs. 
The SNRs belonging to the latter group have relatively small 
radii in general, implying that most of them are likely the remnants of 
SN IIL/b or SN IIn that had strong mass loss before the explosion. 
I also comment on the ``[Fe II]-H$_2$ reversal'' in SNRs 
and on using the [Fe II]-line luminosity as 
an indicator of the supernova (SN) rate in galaxies. 
In the mid- and far-infrared regimes, thermal dust emission is dominant.  
The dust in SNRs can be heated either by collisions with 
gas species in a hot plasma or by radiation from a shock front. 
I discuss the characteristics of the infrared morphology of 
the SNRs interacting with molecular clouds and their dust heating processes. 
Finally, I give a brief summary of the detection 
of SN dust and crystalline silicate dust in SNRs. 

\end{abstract}

\firstsection 

\section{Introduction}

Infrared (IR) covers 3 decade logarithmic scales in wavelength, 
from 1 to 1000 $\mu$m. 
This is the waveband in which we observe emission from dust, 
forbidden fine-structure lines from various metallic atoms and ions,  
molecular lines, and H-recombination lines.
These diverse and unique emission features, together with their 
relatively small extinctions, make the IR band particularly useful 
for studying various physical and astrophysical 
processes related to shocks and supernova remnants (SNRs). 

During the past 10 years, significant progress has been made in the IR  
study of SNRs as a result of space missions equipped with mid- and far-IR
instruments and the development of wide-field IR cameras and broadband 
spectrometers.
In this paper, I shall talk about two particular spectral features
often found in SNRs:
(1) [Fe II] emission lines in the near-IR (NIR) band, which is 
the most prominent NIR spectral feature 
in SNRs, and (2) dust continuum emission in 
mid- and far-IR spectra. 
For the [Fe II] lines, I briefly review the basic physics, 
summarize observational results, and then discuss the characteristics 
of [Fe II]-bright SNRs along with some related issues.
For the dust emission, as there are other papers on this topic in this volume, 
I simply present some recent topics that are relevant to supernovae (SNe) and 
SNR environments.     

\section{NIR [Fe II] Emission from SNRs}
\subsection{NIR [Fe II] Emission Lines and J Shocks}

\begin{figure}[t]
\begin{center}
 \includegraphics[width=5.0in]{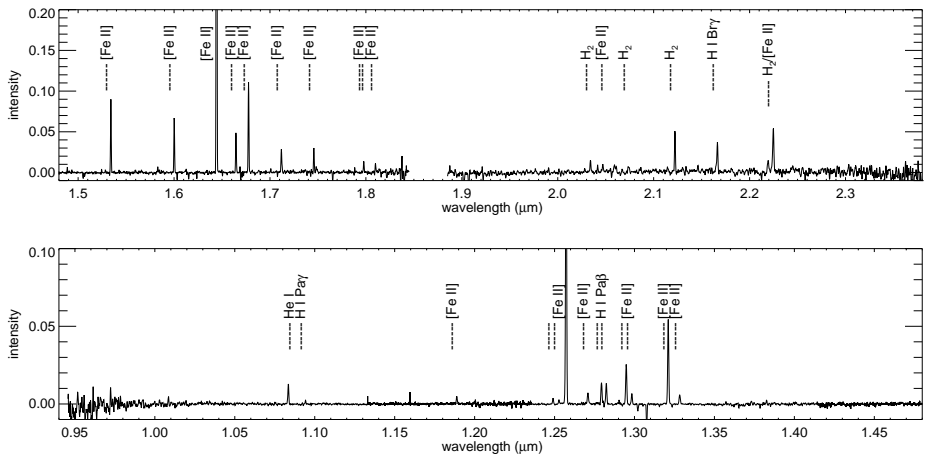}
 \caption{TripleSpec spectrum of the SNR G11.2-0.3, 
showing numerous [Fe II] lines and other emission 
lines (Courtesy of D.-S. Moon). 
The extinction to the source is large ($A_V=16$~mag), so that the 
observed line intensities can be significantly 
different from the intrinsic ones.
}
\label{fig1}
\end{center}
\end{figure}

In the NIR spectra of SNRs, 
[Fe II] emission lines are usually the most prominent 
unless there is heavy-element-enriched SN ejecta (Fig. 1). 
This contrasts with photoionized HII regions where 
H recombination lines are much stronger;
i.e., [Fe II] 1.257~$\mu$m/Pa$\beta$
 ($\sim 3\times $ [Fe II] 1.644~$\mu$m/Pa$\alpha$)=2--8 in SNRs
whereas it is 0.013 in Orion 
(\cite[Oliva et al. 1989]{oliva89}; 
\cite[Mouri et al. 2000]{mouri00}). 
Such a large difference arises because Fe atoms
in photoionzed gas are in higher ionization stages and 
also probably because the Fe abundance in shocked gas is enhanced by dust 
destruction. Therefore, [Fe II] emission 
can be used as a tracer of fast radiative atomic shocks, 
although strong [Fe II] lines may be observable 
in sources ionized by X-rays, e.g., 
in active galactic nuclei (\cite[Mouri et al. 2000]{mouri00}).

The Fe$^+$ ion has four ground terms, each of which has
3--5 closely-spaced levels to form a 16 level system
(\cite[Pradhan \& Nahar 2011]{pradhan11}).
The energy gap between the ground level and the excited levels
is less than $1.3\times 10^4$~K, and thus, 
these levels are easily excited in the postshock cooling region.
The emission lines resulting from the
transitions among these levels appear in the visible to far-infrared bands
(Fig. 2 right).
In the NIR JHK bands, 10--20 [Fe II] lines are visible; these include   
the two strongest lines at 1.257 and 1.644 $\mu$m.
The ratios of these lines provide a very good density
diagnostic and an accurate measure of extinction to the emitting region.

\begin{figure}[t]
\begin{minipage}{.5\textwidth}
\hspace*{-0.5 cm}
\includegraphics[width=3.3in]{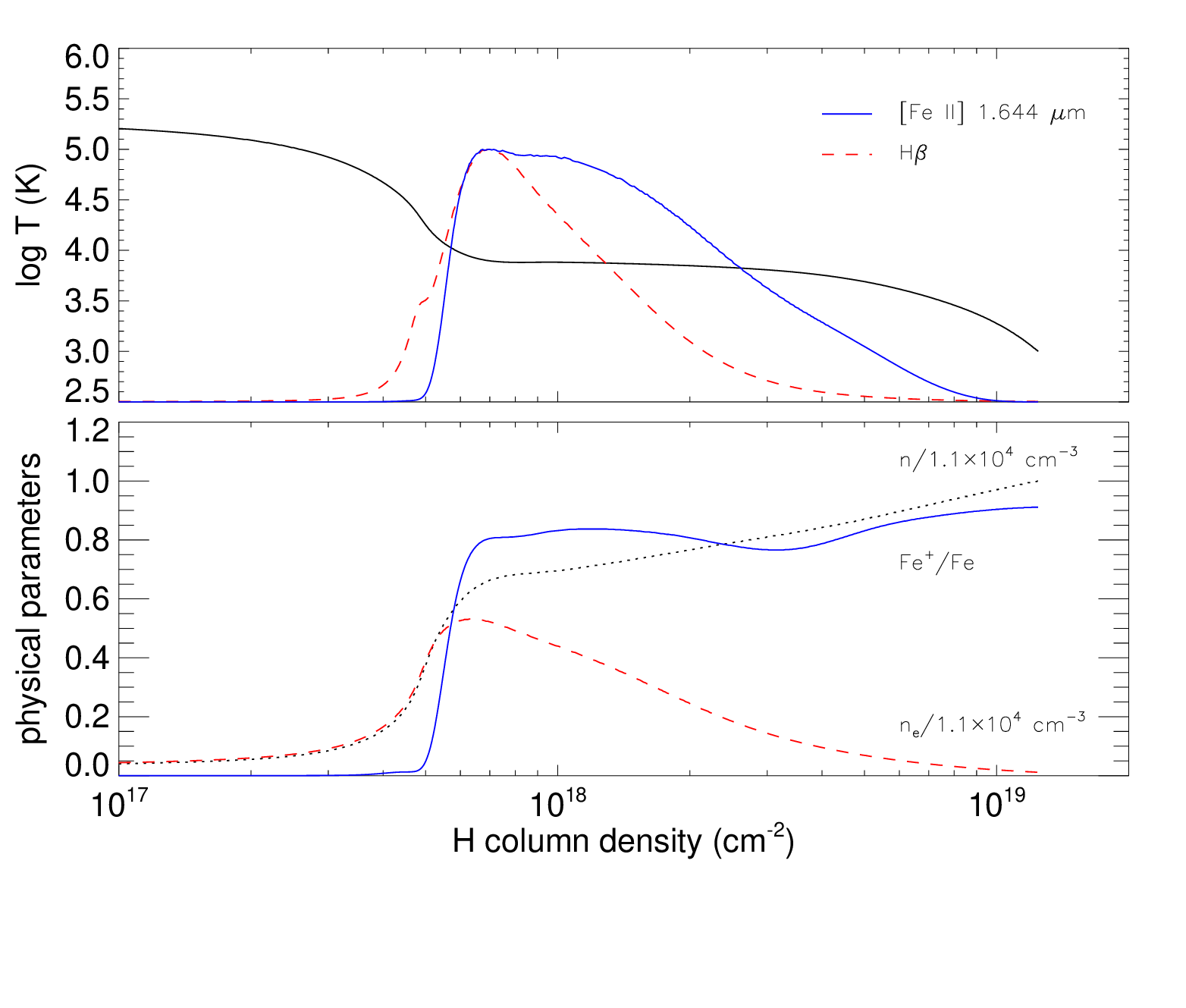}
\end{minipage}
\begin{minipage}{.5\textwidth}
\hspace*{0.5 cm}
\vspace*{-1.8 cm}
\includegraphics[width=2.4in]{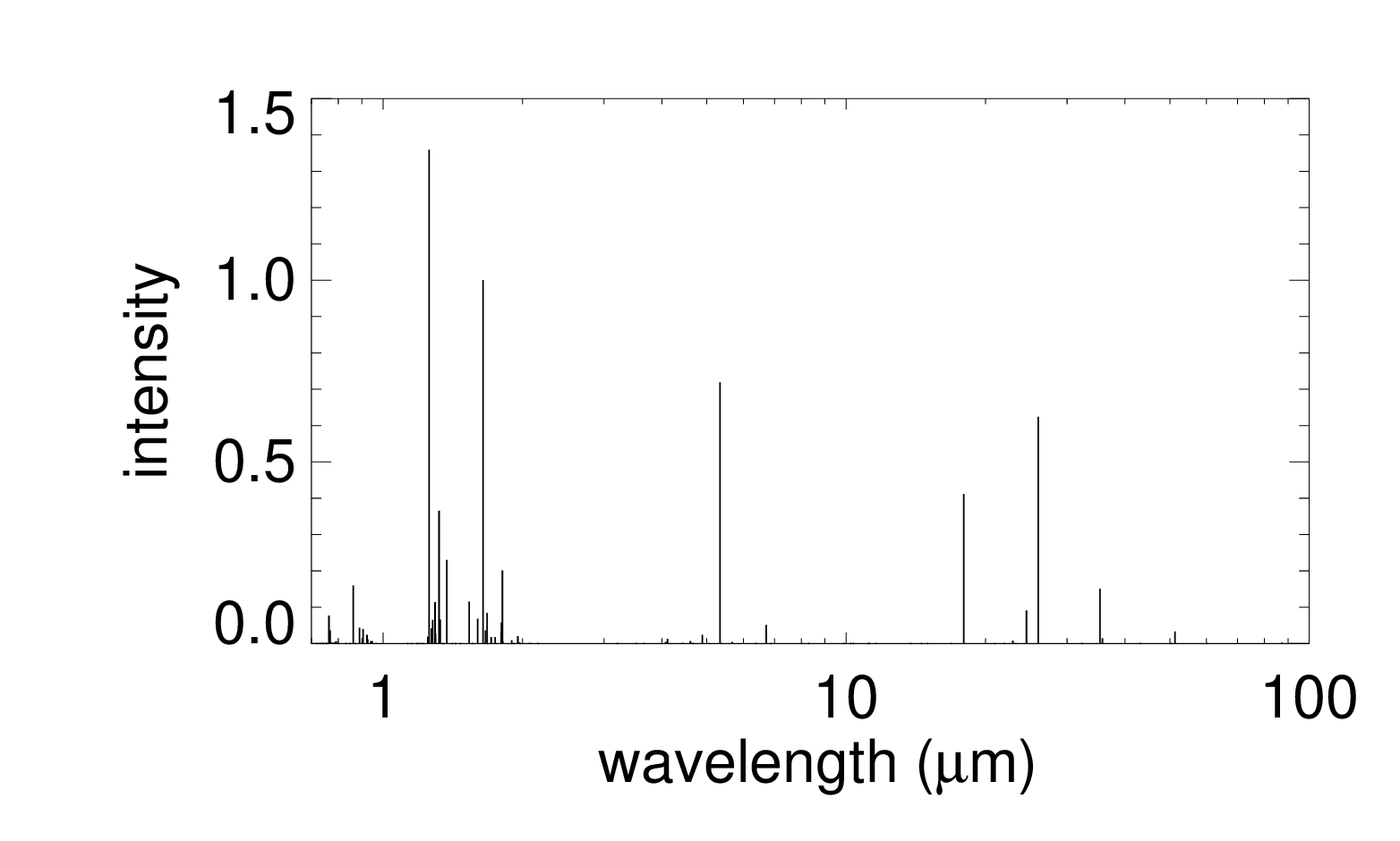}
\end{minipage}
\vspace*{-1.0 cm}
\caption{(Top left) Temperature profile as a function of swept-up H-nuclei
column density for a 150~\kms\ shock propagating into an ambient medium 
of $n_0=100$~cm$^{-3}$ and $B_0=10$~$\mu$G. 
[Fe II] 1.644 $\mu$m and H$\beta$ 
line emissivities are overplotted in an arbitrary linear scale.
(Bottom left) Profiles of H nuclei density ($n$),
electron density ($n_e$) and fraction of Fe in Fe$^+$ (Fe$^+$/Fe) for the same shock.
(Right) Synthesized IR spectrum of [Fe II] lines from the shock,
normalized to the [Fe II] 1.644 $\mu$m line intensity. 
The calculation is done by using the Raymond code.
}
\label{fig2}
\end{figure}

[Fe II] emission lines in SNRs are mostly emitted from cooling gas behind
radiative atomic shocks.
Figure 2 (top left) shows the temperature profile
in the postshock cooling layer of a radiative shock.
At $N_{\rm H}\sim 5\times 10^{17}$~cm$^{-2}$
the cooling becomes important and
the temperature abruptly drops to $\sim 8,000$ K.
Then the temperature remains constant over an extended region,  
where the heating is maintained by UV radiation
generated from the hot gas immediately behind the shock front.
The corresponding profiles of H nuclei and electron densities
are shown in the bottom left frame of Figure 2, together with the 
Fe$^+$ fraction profile. 
Note that, since the ionization potential of the iron atom is 7.90 eV,
far-UV photons from the hot shocked gas
can penetrate far downstream to
maintain the ionization state of Fe$^+$
where H atoms are primarily neutral.
Most of the [Fe II] emission, however, originates from the
temperature plateau region where the ionization fraction is
not too low, as shown in Figure 2. 
Numerical shock models covering some parameter spaces are available in 
\cite[Hollenbach et al. (1989)]{hollenbach89b},
\cite[Mouri et al. (2000)]{mouri00}, and \cite[Allen et al. (2008)]{allen08}.
A grid of shock models with updated atomic parameters is 
in preparation by the author of this paper.

\subsection{NIR [Fe II] Observations of SNRs}

The first detection of the [Fe II] 1.644 $\mu$m line in an SNR
was reported by \cite[Seward et al. (1983)]{seward83} on MSH 15$-$52. 
After that, about a dozen Galactic and LMC SNRs have been observed 
in NIR [Fe II] lines.
This number will increase with the completion 
of the UWIFE (UKIRT Wide-field Infrared Survey for Fe$^+$) project, 
which is an 
unbiased survey of the [Fe II] 1.644 $\mu$m line of the 
inner Galactic plane ($\ell=7^\circ$ to 65$^\circ$;  
$|b|\le 1.3^\circ$) 
using the UKIRT 4-m telescope. The UWIFE project is a
``cousin'' of the UWISH2 project,  
which covers the same area in the H$_2$ $v=1\rightarrow 0$ S(1) line
at 2.122 $\mu$m (\cite[Froebrich et al. 2011]{froebrich11}). 
Lee, Y.-H. et al. (this volume) introduce the two projects and 
present preliminary results on SNRs.
In short, there are 77 SNRs in this area, and 
about 20\%--30\% of them are detected in [Fe II] and/or H$_2$ lines,
more than half of which are new detections.

The SNRs bright in [Fe II] emission lines may be divided into two groups: 
(1) middle-aged SNRs interacting with dense molecular (or atomic) clouds, and 
(2) young SNRs interacting with the dense circumstellar medium (CSM). 

{\underline{\it Middle-aged SNRs bright in [Fe II] emission}}.
Prototypical SNRs belonging to this category are  
W44, 3C391, and IC 443. All of them are interacting with 
molecular clouds (MCs). An indication of the MC interaction 
in the NIR band is the presence of H$_2$ ro-vibrational emission lines that arise
from slow, non-dissociative C shocks 
propagating into dense molecular gas of low-fractional ionization. 
Therefore, these middle-aged SNRs
that are bright in [Fe II] emission are also bright in 
H$_2$ emission (Fig. 3). 		

\begin{figure}[t]
\begin{minipage}{.5\textwidth}
\hspace*{0.4 cm}
\includegraphics[width=3.0in]{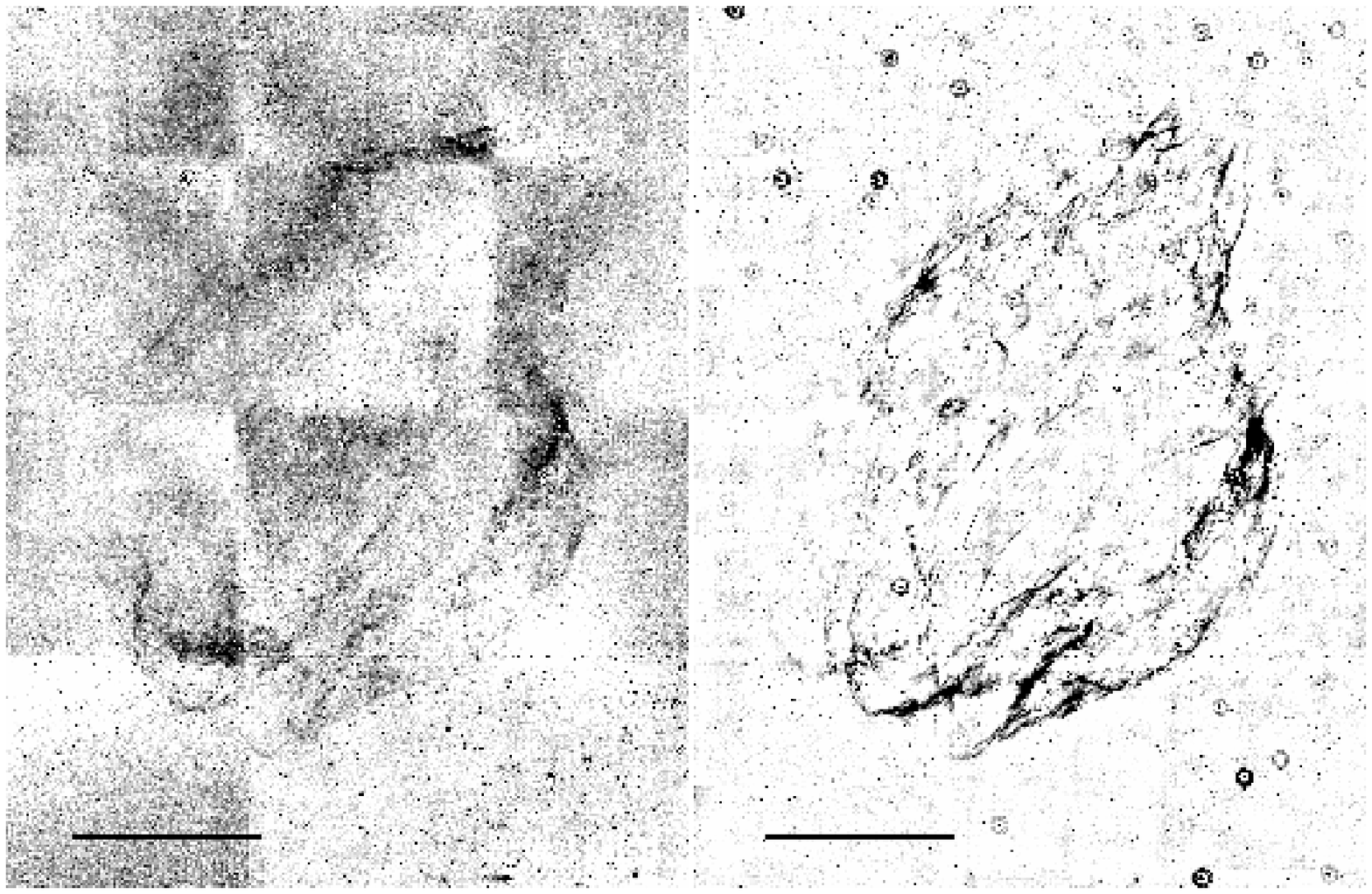}
\end{minipage}
\begin{minipage}{.5\textwidth}
\hspace*{2.0 cm}
\includegraphics[width=1.5in]{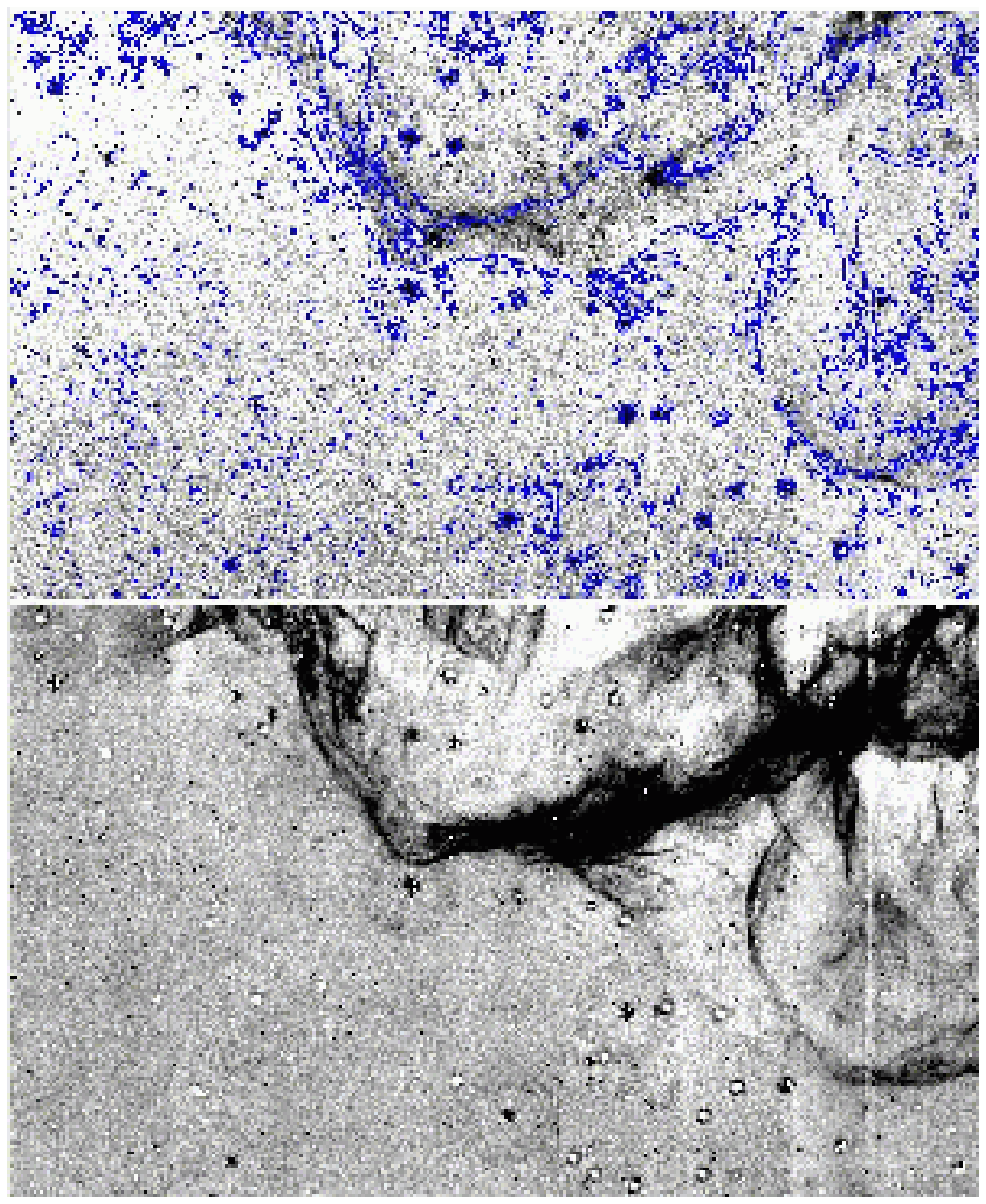}
\end{minipage}
 \caption{UWIFE [Fe II] 1.644 $\mu$m (left) and 
UWISH2 H$_2$ 2.122 $\mu$m (middle) images of W44. 
The scale bar corresponds to 10$'$.
The images on the right show zoomed-in views of the 
southern area in the [Fe II] (top) and H$_2$ (bottom) 
emissions, respectively.
The contours of the H$_2$ emission are overlaid 
on the top right [Fe II] image.
}
\label{fig3}
\end{figure}

Figure 3 shows [Fe II] 1.644 $\mu$m and H$_2$ 2.122 $\mu$m line 
images of W44; we see 
that the overall morphologies of the SNR in the two lines are 
similar, although 
the former appears rather diffuse and confined to the SNR boundary,  
whereas the latter is considerably filamentary and fills the 
entire SNR (see also \cite[Reach et al. 2005]{reach05}). 
A detailed inspection shows that, in some areas, there is a good
spatial correlation between the two, whereas in other areas,
the correlation is less significant.
The detection of 
both [Fe II] and H$_2$ emission lines is consistent with 
the general consensus that 
a molecular cloud is clumpy, being composed of 
dense clumps embedded in a rather diffuse interclump gas. 
An SNR produced inside a MC expands into 
the interclump medium while engulfing dense clumps. 
In late stages, the SNR shock in the interclump medium 
becomes radiative, so that the SNR becomes  
surrounded by a fast-expanding ($\sim 100$~\kms) 
atomic shell, which is observable in \schi 21-cm line
(e.g., \cite[Koo \& Heiles 1995]{koo95}; 
\cite[Chevalier 1999]{chevalier1999}). 
The [Fe II] emission in such SNRs can originate in two regions; 
either from radiative atomic shocks in the interclump medium or 
from reflected shocks in the SNR shell 
generated by the interaction with molecular clumps.
A detailed, comparative study of [Fe II] and H$_2$ emissions 
should reveal more detailed information about the structure of molecular clouds.

\begin{figure}[t]
\begin{minipage}{.5\textwidth}
\hspace*{-1.0 cm}
\includegraphics[width=4.0in]{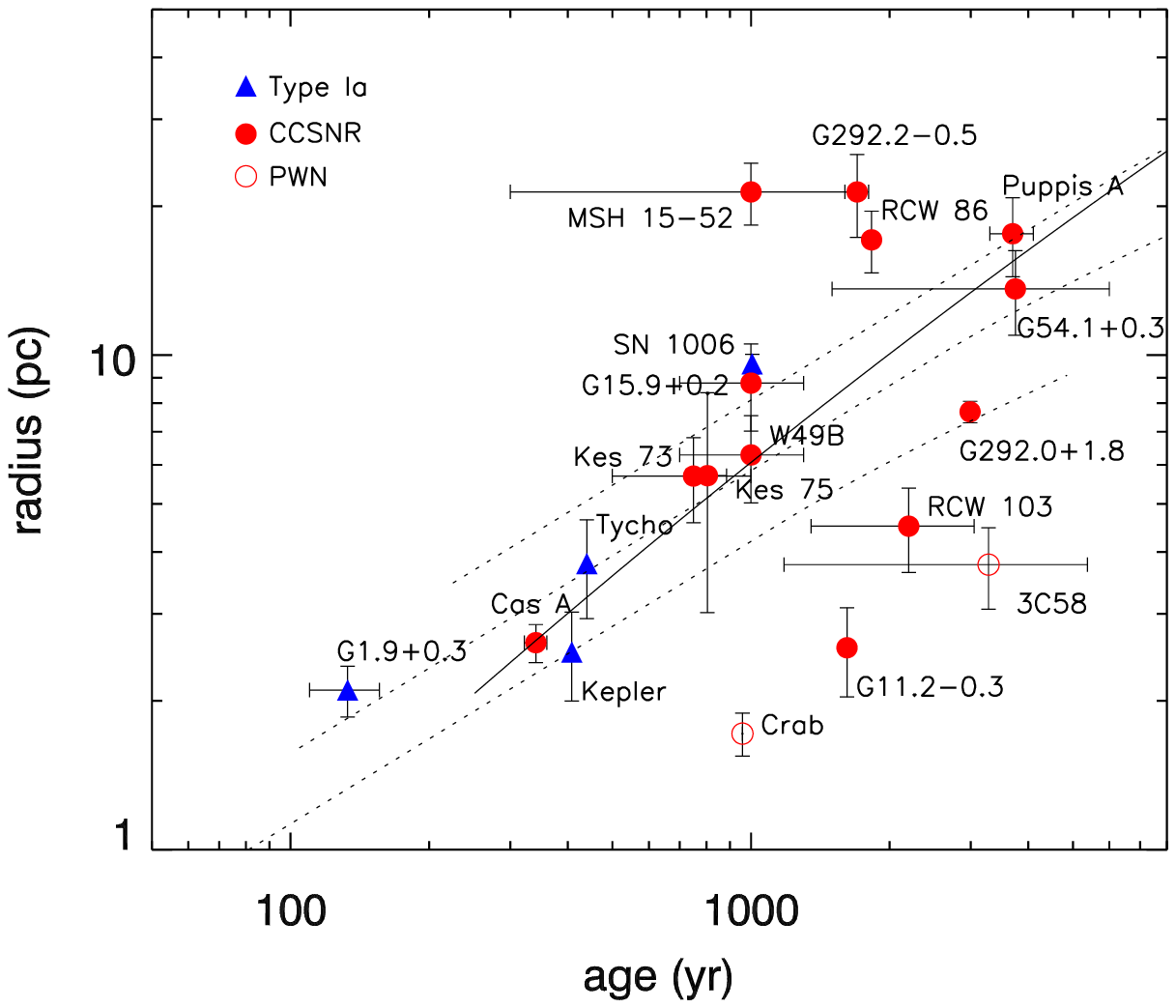}
\end{minipage}
\begin{minipage}{.5\textwidth}
\hspace*{3.0 cm}
\vspace*{0.7 cm}
\includegraphics[width=1.3in]{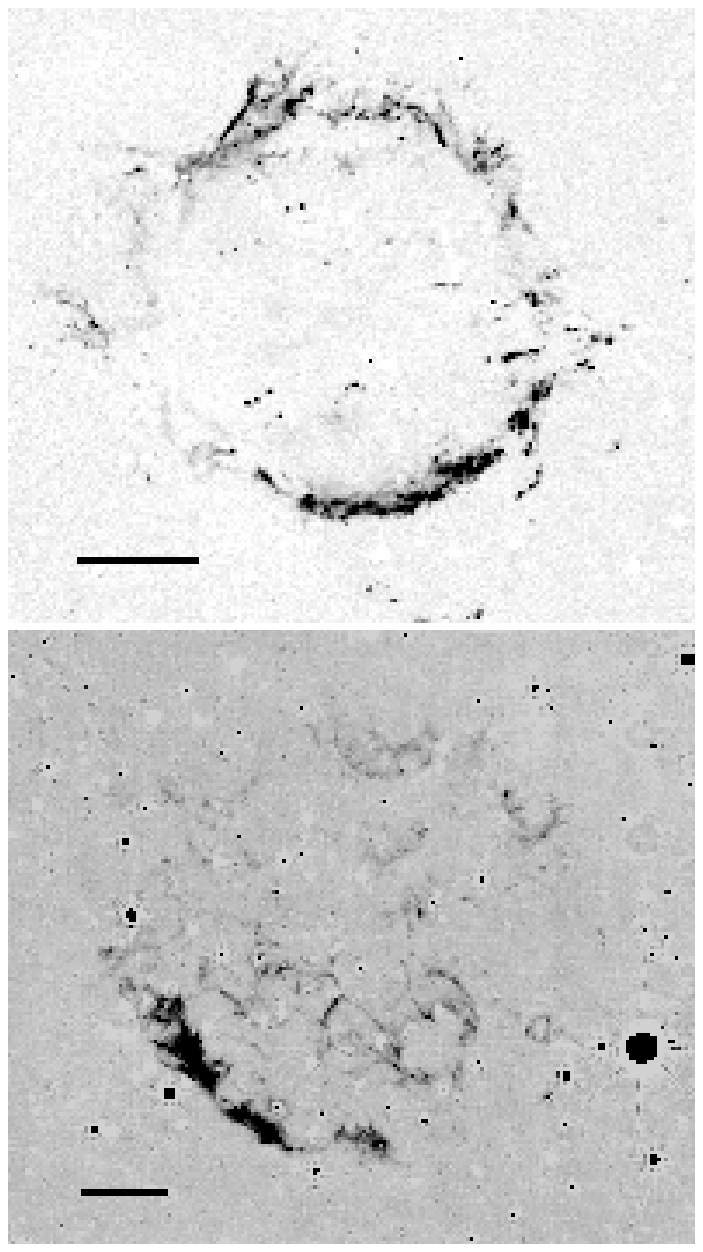}
\end{minipage}
\vspace*{-0.3 cm}
 \caption{(Left) Radius versus age of young SNRs.
The dotted lines are models for SNRs in uniform ambient media of 
$n_0=1$, 0.1, and 0.01 cm$^{-3}$, respectively
(\cite[Truelove \& McKee 1999]{truelove1999}; $n=7$ ejecta model with 
$M_{\rm ej}=5~M_\odot$ and $E_{\rm SN}=10^{51}$ ergs).  
The solid line is for an SNR in the RSG wind case
(\cite[Chevalier\& Oishi 2003]{chevalier03}; see text for the 
parameters of the model.)
Note that the ones 
marked by empty circles are pulsar wind nebulae, so they do not represent 
true sizes of SNRs. (Right) [Fe II] images of Cas A and G11.2-0.3 from 
top to bottom. The scale bars correspond to 1 pc at the distances of the SNRs. 
}
\label{fig4}
\end{figure}

{\underline{\it Young SNRs bright in [Fe II] emission}}.
There are also young SNRs bright in [Fe II] emission. 
Prototypical ones are Cas A, G11.2$-$0.3, RCW 103, and W49B, 
all of which are
core-collapse SNRs (CCSNRs) with a central stellar remnant, except 
W49B, where no central source has been detected.
One way to infer the environment of these [Fe II]-bright young SNRs is to 
inspect diagrams such as Figure 4, where we compare the radii and ages of young SNRs. (For a discussion of SN types of young
SNRs, see \cite[Chevalier 2005]{chevalier05}.)
As in the figure, there is a trend of increasing SNR size with age, 
but with a large scatter, indicating    
diverse SN environments and also possibly 
diverse SN explosion energies. 
Among Type Ia SNRs, 
SN 1006 is located about 500 pc above the Galactic plane 
where the ambient density is $\sim 0.05$~cm$^{-3}$, whereas  
the Kepler SNR is interacting with a relatively dense ($\simgt 1$ cm$^{-3}$) medium.
Young CCSNRs are interacting either 
with CSM or a wind bubble created in 
the main-sequence lifetime of their progenitors. 
Cas A, which is Type IIb SN, for example, is  
interacting with a dense red supergiant (RSG) wind.
The solid line is a model for Cas A
from \cite[Chevalier \& Oishi (2003)]{chevalier03}, 
but assuming $\dot M_w=3\times 10^{-5} M_\odot {\rm yr}^{-1}$,
$v_w=15$~\kms, $M_{\rm ej}=5M_\odot$, and $E_{\rm SN}=10^{51}$ ergs where 
$\dot M_w$ is the wind mass-loss rate, 
$v_w$ is the wind speed, $M_{\rm ej}$ is the ejecta mass, and 
$E_{\rm SN}$ is the explosion energy.
Cas A follows this ``Cas A-like'' line 
as long as it continues to interact with the dense CS wind.
There are CCSNRs which fall well below  
the Cas A-like line, i.e., G11.2$-$0.3, RCW 103, and G292.0+1.8. 
Note that the first two SNRs are bright in [Fe II] lines.
(G292.0+1.8 has not been observed in [Fe II] emission.)
Their relatively small sizes could be due to either a small explosion energy 
($< 10^{50}$~ergs) or/and dense CSM.
The strong [Fe II] lines in these SNRs suggest that 
it is more likely because of dense CSM and that the CSM is 
much denser than that of Cas A.
Hence, they are likely the remnants of massive SN IIL/b or SN IIn. 
On the other hand, there are remnants 
much larger than Cas A: MSH 15$-$52, G292.2$-$0.5, and RCW 86.  
These remnants might have exploded inside a large bubble, and 
thus, they are candidates for SN Ib/c, although a 
large SN explosion energy ($> 10^{52}$~ergs) could be another possibility.
Figure 4 suggests that all four [Fe II]-bright young CCSNRs 
mentioned at the beginning of this paragraph are SN IIL/b candidates 
interacting with dense CSM. 
(For W49B, however, the bipolar Type Ib/c SN origin has been suggested.
See \cite[Lopez et al. 2013]{lopez13} and references therein.)

The [Fe II] emission in these young SNRs originate from both  
shocked CSM and shocked SN ejecta.
In Cas A, it is well known that there are two types of 
knots detected in the visible waveband: quasi-stationary flocculi (QSFs),  
which are dense CS knots moving at a few hundred 
kilometers per second,  
and fast moving knots (FMKs),  
which are metal-rich SN ejecta knots moving at several 
thousand kilometers per second. 
Lee, Y.-H. et al. (this volume)  
show that there are also fast-moving [Fe II] knots
that lack other metallic lines, which could be 
pure Fe ejecta synthesized in the innermost SN region. 
In G11.2$-$0.3, which is known as a cousin of Cas A 
because of its similar morphology (Fig. 4), 
the knots in the central area have 
radial velocities of $\sim 1,000$ \kms, 
which suggests that they might be SN ejecta 
(\cite[Moon et al. 2009]{moon09}). 
Again, their spectra do not show metallic lines other than Fe. 
The bright filament in the southeast of G11.2$-$0.3, on the other hand, 
appears to be composed of mostly 
dense CSM.
For the [Fe II] emission features in RCW 103 and W49B,  
detailed spectroscopic studies are yet to be performed 
(cf. \cite[Oliva et al. 1999]{oliva99}; \cite[Keohane et al. 2007]{keohane07}). 

\subsection{Some Issues}

{\underline{\it [Fe II] - H$_2$ reversal }}.
Since the early days of NIR observations of SNRs, 
it has been known that 
there are SNRs with H$_2$ filaments lying beyond [Fe II] 
filaments, i.e., further out from the SNR center,
which is not easily explained by shock models  
(\cite[Graham et al. 1991]{graham91}; 
\cite[Oliva, Moorwood, \& Danziger 1990]{oliva90}; 
\cite[Burton \& Spyromilio 1993]{burton1993}). 
We now have more sources showing similar patterns, e.g., 
G11.2-0.3, W49B, and 3C396 
(\cite[Koo et al. 2007]{koo07}; \cite[Keohane et al. 2007]{keohane07}; 
\cite[Lee et al. 2009]{lee09}).
W44 in Figure 3 is another example.
Hence, we need an explanation for this ``[Fe II]-H$_2$ reversal''; 
i.e., 
we need to know what the exciting mechanisms of the H$_2$ emission is and 
how they excite the H$_2$ gas beyond the SNR. 
Some proposed mechanisms are fluorescent UV excitation, 
X-ray heating, magnetic precursors, and reflected shocks, 
but high-resolution NIR spectroscopic studies are needed 
to address the issue.

{\underline{\it [Fe II] luminosity as an SN rate indicator }}.
Can the [Fe II] 1.257 or 1.644 $\mu$m luminosities  
be used as an indicator of galactic SN rates?
Several groups have addressed this for starburst galaxies 
and have derived a conversion factor of 
{SN~ rate}
=0.03--0.1 $L_{[Fe II]}/10^6 L_\odot$ yr$^{-1}$
where $L_{[Fe II]}$ is the [Fe II] 1.644 $\mu$m luminosity
(\cite[Morel et al. 2002]{morel02}; 
\cite[Alonso-Herrero et al 2003]{alonso03};
\cite[Rosenberg et al. 2012]{rosenberg12}).
In the case of nearby starburst galaxies M82 and NGC 253, 
however, 
70\%--80\% of the [Fe II] emission is known to be diffuse emission of unknown origin, 
although it is speculated to be related to the SN activity therein 
(\cite[Greenhouse et al. 1997]{greenhouse97}; 
\cite[Alonso-Herrero et al. 2003]{alonso03}).
In the Galaxy, 
the [Fe II]-bright SNRs represent  20\%-30\% 
of the {\em  known} $\sim 300$ SNRs, 
which occupy a small fraction of the entire set of SNRs present in the Galaxy. 
It is clear that we need to have better understanding of the population of 
[Fe II]-bright SNRs  
and also the origin of the diffuse [Fe II] emission
to obtain a more reliable relation between 
$L_{[Fe II]}$ and the SN rate in galaxies.

\section{Dust IR Emission from SNRs}

\subsection{Dust Heating in SNRs Interacting with MCs}

\begin{figure}[b]
\begin{minipage}{.5\textwidth}
\hspace*{1.0 cm}
\includegraphics[width=2.0in]{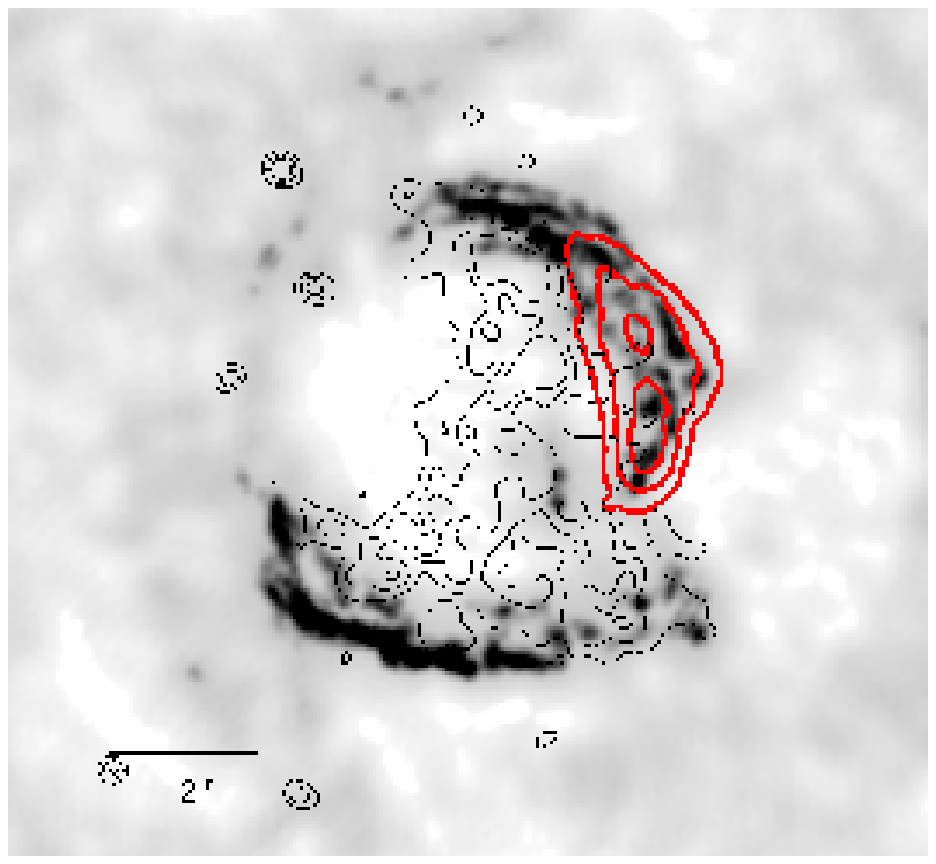}
\end{minipage}
\begin{minipage}{.5\textwidth}
\hspace*{0.5 cm}
\includegraphics[width=2.0in]{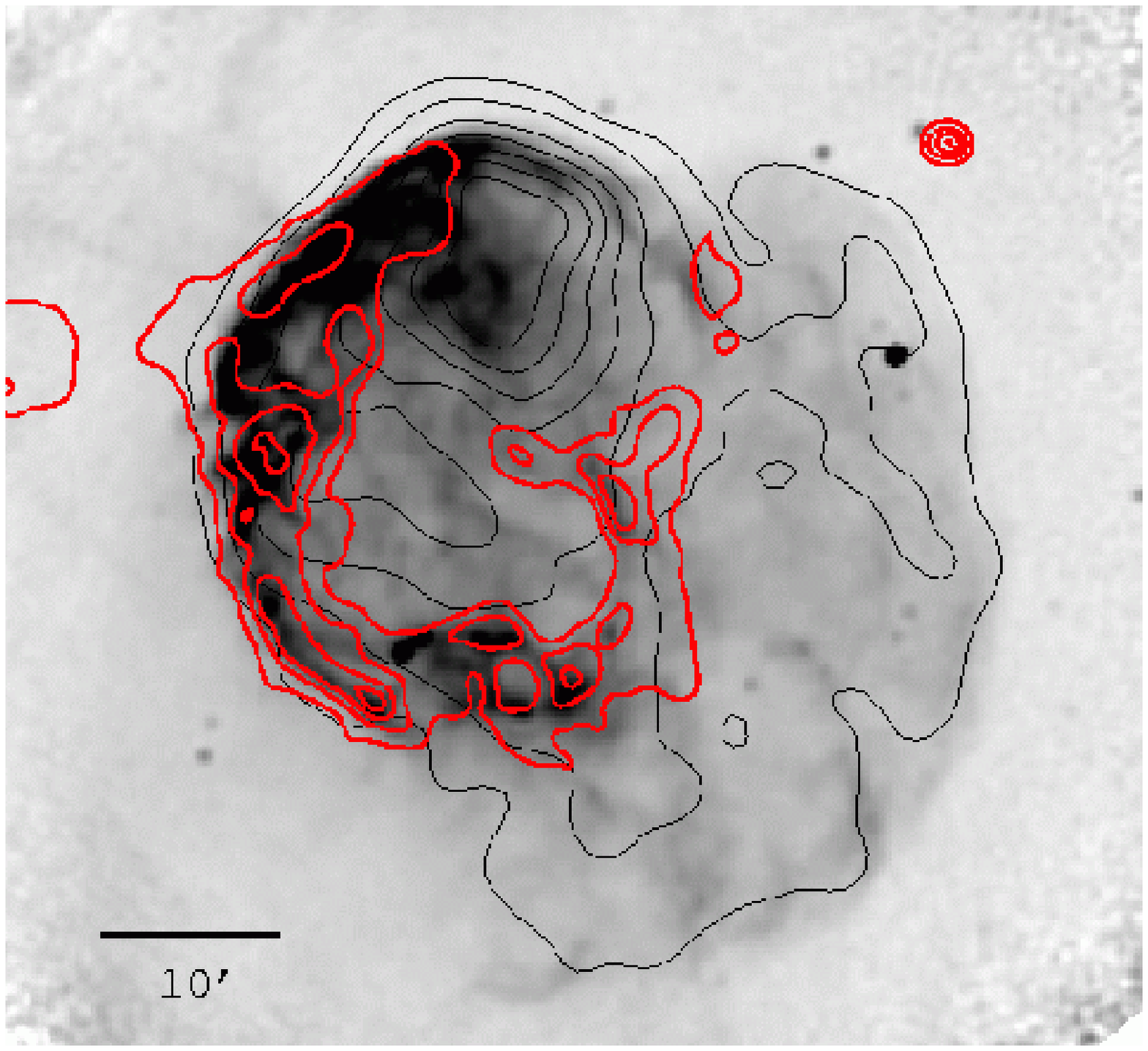}
\end{minipage}
\caption{(Left) ATCA 20-cm image of Kes 17 with overlaid 
XMM 0.2--12 keV X-ray (thin) 
and AKARI 65 $\mu$m (thick) contours. 
(See also \cite[Lee et al. 2011]{lee11}.)
(Right) VLA 20-cm image of IC 443 with overlaid 
ROSAT 0.2--2.4 keV X-ray (thin) and AKARI 90 $\mu$m (thick) contours.
}
   \label{fig5}
\end{figure}

Dust in SNRs can be heated either collisionally or radiatively.
In SNRs with fast, non-radiative shocks, dust grains are heated by collisions 
with gas particles, mainly electrons, in a hot plasma behind 
the shocks (e.g., \cite[Dwek et al. 2008]{dwek08}). 
Many SNRs have mid- and far-IR morphology almost identical 
to that of X-ray, which suggests that the IR emission 
in these SNRs is primarily from collisionally-heated dust grains. 

SNRs interacting with MCs generally have IR morphology  
different from the X-ray morphology; they appear shell-like in 
IR whereas they are centrally brightened in X-rays (Fig. 5). 
The dust in these SNRs cannot be collisionally-heated by 
X-ray emitting gas. Instead the likely source of the heating is 
radiation from the shock front. 
In radiative shocks, the UV radiation from the cooling postshock
gas could be much stronger than the general interstellar radiation 
field (e.g., \cite[McKee et al. 1987]{mckee87}).
This UV radiation, dominated by trapped 
Ly$\alpha$ photons, heat the dust in the cooling layer, 
and, subsequently,  
the infrared radiation from these hot dust grains 
heats the dust at larger column densities 
(e.g., \cite[Hollenbach et al. 1979]{hollenbach79}).
The far-IR bright regions in MC-interacting SNRs (Fig. 5)
are probably where the radiation field is strong 
and the ambient density is high.
\cite[Andersen et al. (2011)]{andersen11} carried out a systematic 
study of the dust emission from MC-interacting SNRs found in the 
GLIMPSE survey, and derived dust temperatures of 29--66 K 
from the Spitzer MIPS spectral energy distribution (60--90~$\mu$m).
There could be a dust component at a lower temperature, however,
because the SNRs interacting with MCs are bright in the 
far-IR waveband beyond the MIPS coverage
(e.g., see \cite[Lee et al. 2011]{lee11}). 
A systematic study of MC-interacting SNRs, including the AKARI and Herschel far-IR data, 
will be useful to understand the heating mechanisms and also the processing of dust grains 
in these SNRs.

\subsection{Star Dust in SNRs}

{\underline{\it SN dust in young CCSNRs}}. 
The dense, metal-rich, cooling
SN ejecta can effectively provide an environment for dust to condense.
In the high-redshift galaxies, where low-mass stars
do not have enough time to evolve to AGB stars, SNe could be the
main contributors of dust, depending on the dust yield
(\cite[Dwek \& Cherchneff 2011]{dwek11}).
Theoretical studies have shown that
as much as 1 $M_\odot$ of different dust species 
can form in SN IIP with massive H envelopes, 
whereas in SN IIL/b or SN Ia with little or no H envelopes
only a limited amount of dust could form
(\cite[Nozawa et al. 2010]{nozawa10}; \cite[Nozawa et al. 2011]{nozawa11}).

In observational studies, however, only a
very small amount of dust in SNe has been detected, i.e.,
$\simlt 10^{-3} M_\odot$.
(see \cite[Gall et al. 2011]{gall2011} and references therein).
It is only toward the LMC SN 1987A and some young Galactic SNRs
where a significant amount of SN dust has been detected:
In 1987A, \cite[Matsuura et al. (2011)]{matsuura11} reported
detection of 0.4--0.7~$M_\odot$ of dust, whereas, 
in the Galaxy, 0.1--0.2 $M_\odot$ of dust has been
detected in Cas A and the Crab nebula which are SNIIb and SN IIP(?), respectively.
In another SN IIP candidate, G54.1+0.3, a dust ring of
0.58--0.86 $M_\odot$ has been detected around its pulsar wind nebula,
but the nature of the ring is not yet conclusively identified 
(\cite[Koo 2012]{koo12}).

{\underline{\it Crystalline silicate dust in MSH 15$-$52}.
Essentially all dust grains in the ISM are amorphous.
Crystalline silicate dust grains have been found mainly in 
evolved stars and young stellar objects, indicating that 
they form in situ in circumstellar disks and/or outflows of 
these objects (\cite[Henning 2010]{henning10}). 
In this regard, the detection of crystalline silicates 
in the SNR MSH 15$-$52 is interesting (\cite[Koo et al. 2011]{koo11}).

As we mentioned in \S~2.2, MSH 15$-$52, is a 
young ($\sim 1,000$~yr) SNR probably expanding inside a bubble, 
suggesting progenitor SN type of Ib/c (see Fig. 4).
The remnant has a central pulsar, and there is 
an O star (Muzzio 10) and a bright MIR source (IRAS 15099$-$5856) 
lying very close from the pulsar, i.e., 
$18''$ and $31''$ (or 0.35 pc and 0.60 pc at 4 kpc) to north, 
respectively.
IRAS 15099$-$5856 is probably a dusty cloud heated by Muzzio 10.
What is special to this mid-IR source is that it has prominent crystalline 
silicate spectral features. 
\cite[Koo et al. (2011)]{koo11} proposed a scenario 
where the SN progenitor and Muzzio 10 were in a binary system and 
IRAS 15099$-$5856 is a CSM survived from the SN blast wave 
due to the shielding by Muzzio 10 
(see \cite[Koo 2012]{koo12} for more details). 
MSH 15$-$52 appears to be the first case in which 
crystalline silicates have been observed to be associated with a SNR. 

\noindent
{\bf Acknowledgements}

\noindent
I wish to thank Lee, Y.-H., Jeong, I.-G, and Moon, D.-S. 
for their help with figures.
My research is supported by Basic Science
Research Program through the National Research Foundation of Korea (NRF) funded by
the Ministry of Education, Science and Technology (NRF-2011-0007223).

\begin{discussion}

\end{discussion}

\end{document}